\documentclass[12pt]{report}
\usepackage[dvips]{graphicx}
\newcommand{\sptwo}{1.4}

\newcommand{\doublespace}{\edef\baselinestretch{\sptwo}\Large\normalsize}

\begin{document}
\doublespace
\begin{center}
{\bf Dynamics of  Strongly Interacting Fermi Gases of Atoms in a Harmonic Trap}\\
\renewcommand\thefootnote{\fnsymbol{footnote}}
{Yeong E. Kim \footnote{ e-mail:yekim$@$physics.purdue.edu} and
Alexander L. Zubarev\footnote{ e-mail: zubareva$@$physics.purdue.edu}}\\
Purdue Nuclear and Many-Body Theory Group (PNMBTG)\\
Department of Physics, Purdue University\\
West Lafayette, Indiana  47907\\
\end{center}

\begin{quote}
Dynamics of strongly interacting trapped dilute Fermi gases 
 is investigated
 at zero temperature.
 As an example of application
 we consider the expansion of the cloud of fermions
initially confined in an anisotropic harmonic trap, and  study 
the equation of state dependence of the radii of the trapped cloud and 
the collective
oscillations in the vicinity of a Feshbach resonance.

\end{quote}

\vspace{5mm}

\vspace{55 mm}
\noindent

\pagebreak

The newly created ultracold trapped Fermi gases with tunable atomic scattering
 length [1-10] in the vicinity of  a Feshbach resonance offer the possibility to
 study highly
 correlated many-body systems including the cross-over from the
 Bardeen-Cooper-Schrieffer (BCS) phase to the Bose-Einstein condensate (BEC)
 of molecules [1,11-15].

 In this letter we report our investigation of the dynamics of the  strongly
 interacting dilute
Fermi gas (dilute in the sense that the range of interatomic potential is 
small compared with inter-particle spacing) at zero temperature. 
As an example of application we consider the expansion of the cloud of 
$^6Li$ atoms initially confined in an anisotropic harmonic trap,  study
the equation of state dependence of the radii of the trapped cloud and  the
 collective 
oscillations near a broad  Feshbach  resonance at a magnetic field $B= 820 \pm 3$G [16-18].

We consider a Fermi gas comprising a 50-50 mixture of two different states 
confined in a harmonic trap $V_{ext}(\vec{r})=(m/2)(\omega_{\perp}^2 (x^2+y^2)+
\omega_z^2 z^2)$. The s-wave scattering length between the
 two fermionic species is negative, $a<0$.

Our starting point is the single equation approach  to the time-dependent 
density-functional theory [19]. The basic of this strategy is to 
construct the following  equation 
 $$
i\hbar \frac{\partial \Psi}{\partial t}=-\frac{ \hbar^2}{2 m} \nabla^2 \Psi
+V_{ext} \Psi+V_{xc}\Psi
\eqno{(1)}
$$
that yields the
same $n(\vec{r},t)=\mid \Psi(\vec{r},t) \mid^2$ as the original 
many-fermions
 system.
The dynamics of the
system
 is controlled by an effective single-particle potential $V_{xc}(\vec{r},t)$.
The central problem is the approximation for the xc potential. This is in
general
 a nonlocal functional of the density that depends on the history of the system
(on the behavior of the density at times $t^{\prime}<t$).

The simplest  approximation is to ignore nonlocality in
 space and retardation in time. This leads to the adiabatic local density       
approximation
$$
V_{xc}(\vec{r},t)=[\frac{\partial n \epsilon(
n)}{\partial n}]_{n=n(\vec{r},t)}
\eqno{(2)}
$$
where $\epsilon$
is the ground state energy per particle of the homogeneous system and $n$ is the
 density.
The right hand side of Eq.(2) is the local density approximation for the 
ground
-
state xc potential, but it evaluated at the time-dependent density.
Notice that in the above equation $n$ is the total density of the gas given by the sum of the two spin component.

The adiabatic local density approximation is a remarkably good approximation if the energy gap is much larger
 than the oscillator energies $\hbar \omega_z$, $\hbar \omega_{\perp}$ [20,21].
It is expected that this condition is satisfied for small temperature [20,21].
Here we notice Refs.[22-24] who argue that the ground state of the mixture of
 two species of fermions with different densities (mass) contains both a superfluid 
and a normal Fermi liquid. We do not consider this asymmetrical mixture in the letter.

The ground state energy per particle, $\epsilon(n)$, in the low-density regime,
$k_F\mid a \mid \ll 1$, can be calculated using an expansion in power of
 $k_F \mid a \mid$ 
$$
\epsilon(n)=2 E_F(\frac{3}{10}-
\frac{1}{3 \pi} k_F \mid a \mid+0.055661 (k_F\mid a \mid)^2
-0.00914 (k_F\mid a \mid)^3-
0.018604 (k_F\mid a \mid)^4+...),
\eqno{(3)}
$$
where $E_F=\frac{\hbar^2 k_F^2}{2 m}$ and $k_F=(3 \pi^2 n)^{1/3}$.
The expansion (3) is valid for $3D$. For the case of dimensions $d<3$,
it is known that the quantum-mechanical two-body $t$-matrix vanishes at low
 energies [25].
 The first term in Eq.(3) is 
the Fermi kinetic energy, the second term corresponds to the mean-field
 prediction [26], the next two terms were first  considered in Refs.[27,28] and
 Ref.[29], 
respectively. 

In the $a\rightarrow - \infty$ limit (the Bertsch many-body problem, quoted in 
Ref.[30]) $\epsilon(n)$ is proportional to that of the non-interacting Fermi
 gas
$$
\epsilon(n)=(1+\beta)\frac{3}{10} \frac{\hbar^2 k_F^2}{m},
\eqno{(4)}
$$
where a  universal parameter  $\beta$ [9] is negative and 
$\mid \beta \mid<1$ [30-32].

We also consider the following approximations for $\epsilon (n)$:
$$
\epsilon(n)=E_F (\frac{3}{5}-\frac{(2/(3 \pi)) k_F \mid a \mid}
{1+(6/(35 \pi)) (11-2\ln2) k_F \mid a \mid}),
\eqno{(5)}
$$
and
$$
\epsilon(n)=E_F (\frac{3}{5}-2\frac{\delta_1 k_F \mid a \mid+\delta_2 (k_F \mid a \mid)^2}{1+\delta_3 k_F \mid a \mid+\delta_4 ( k_F \mid a \mid)^2}),
\eqno{(6)}
$$
where $\delta_1=0.106103$, $\delta_2=0.187515$, $\delta_3=2.29188$,
$\delta_4=1.11616$.

While Eq.(5) [31] reproduces first three terms of expansion (3) in low-density
 regime and approximately valid in unitary limit, $\beta=-0.67$,
Eq.(6) reproduces first four terms of expansion (3) in low-density regime and
in unitary limit, $k_F  a  \rightarrow -\infty$, reproduces 
exactly results of
 the recent Monte Carlo calculations [32], $\beta=-0.56$.

It can be proved [33] that every solution of the equation 
$$
i\hbar \frac{\partial \Psi}{\partial t}=-\frac{ \hbar^2}{2 m} \nabla^2 \Psi
+V_{ext} \Psi+\frac{\partial( n \epsilon(n))}{\partial n}\Psi,
\eqno{(7)}
$$
is a stationary point of an action corresponding to the Lagrangian density
$$
\mathcal{L}_0=\frac{i\hbar}{2}(\Psi\frac{\partial\Psi^{\ast}}{\partial t}-
\Psi^{\ast}\frac{\partial\Psi}{\partial t})+\frac{\hbar^2}{2m}\mid \nabla \Psi
\mid^2+\epsilon(n)n+V_{ext}n,
\eqno{(8)}
$$
which for $\Psi=e^{i \phi(\vec{r},t)} n^{1/2}(\vec{r},t)$ can be rewritten as
$$
\mathcal{L}_0=\hbar \dot{\phi}n+\frac{\hbar^2}{2m} (\nabla \sqrt{n})^2+
\frac{\hbar^2}{2m}n (\nabla \phi)^2+\epsilon(n)n+V_{ext}n.
\eqno{(9)}
$$
 The only difference from equations holding for bosons [33,34] is given by
density dependence of $\epsilon(n)$. We do not consider three-body
 recombinations, since these processes play an important role near p-wave
two-body Feshbach resonance [35].

Let us first discuss the expansion of the fermionic superfluid in the $a\rightarrow -\infty$ limit, Eq.(4).
In the hydrodynamic approximation (neglecting quantum pressure term,
$\frac{\hbar^2}{2m} (\nabla \sqrt{n})^2$,
 in Eq. (9)) the corresponding Euler-Lagrange equation admit the simple scaling solution, $
n(\vec{r},t)= n_0(x_i/b_i(t))$ [20].
We note here that the hydrodynamic behavior of a cold Fermi gas [9] is not in
 general direct experimental evidence for superfluidity [36-38].

We take into account the quantum pressure by finding the optimal ground state
energy [39]
$$
\frac{E_0}{N}=\max_{\gamma_x,\gamma_y,\gamma_z}[\sum_{i=1}^3\frac{\hbar \omega_i}{2}\sqrt{\gamma_i}+\frac{3^{4/3}}{4} (1+\beta)^{1/2} N^{1/3}
\prod_{i=1}^3 (\sqrt{1-\gamma_i}\omega_i)^{1/3}].
\eqno{(10)}
$$
In this case the scaling parameters obey the following equations
$$\ddot{b}_i-\frac{\omega_i^2 \gamma_i}{b_i^3}=\frac{\omega_i^2 (1-\gamma_i)}
{b_i \prod_{i=1}^3 b_i^{2/3}},
\eqno{(11)}
$$
at $t=0$ $b_i(0)=1$ and $\dot{b}_i(0)=0$.

The predictions of Eqs.(11) for aspect ratio, $\omega_z \sqrt{1-\gamma_z} b_{\perp}/(\omega_{\perp} \sqrt{1-\gamma_{\perp}} b_z(t))$, are reported in Fig.1 and
 show that the effect of inclusion of the quantum pressure term on the
 expansion
 of superfluid is about 1\%.
 For the reminder of this letter we will use the hydrodynamic approximation.

Now we consider a  general time-dependent harmonic trap, $V_{ext}(\vec{r},t)=
(m/2)\sum_{i=1}^3 \omega_i^2(t) x_i^2$, and  a general $\epsilon(n)$.
A suitable trial function can be taken as
$\phi(\vec{r},t)=\chi(t)+(m/(2 \hbar) \sum_{i=1}^3 \eta_i(t) x_i^2$, 
$n(\vec{r},t)=n_0(x_i/b_i(t))/\prod_jb_j$.
With this ansatz, the Hamilton principle, $\delta\int dt\int\mathcal{L}_0 d^3 r=0$, gives the following equations for the scaling parameters $b_i$
$$
\ddot{b}_i+\omega_i^2(t) b_i-\frac{\omega_i^2}{b_i} \frac
{\int[n^2d\epsilon(n)/dn]_{n=n_0(\vec{r})/\prod_jb_j} d^3 r}
{\int[n^2d\epsilon(n)/dn]_{n=n_0(\vec{r})} d^3 r}\prod_j b_j=0,
\eqno{(12)}
$$
where $b_i(0)=1$, $\dot{b}_i(0)=0$ and $\omega_i=\omega_i(0)$ fix the initial 
configuration of the system, corresponding to the density $n_0(\vec{r})$.

The release energy which corresponds to an integral of motion of Eq.(12) is expressed by
$$
E_{rel}=\frac{1}{N}[\frac{1}{2} \frac{\dot{b}_i^2}{\omega_i^2}
\int n_0^2(\vec{r})\frac{d\epsilon(n_0)}{dn_0}d^3r+\int n_0 \epsilon(n_0/\prod_jb_j)d^3r],
\eqno{(13)}
$$
and for the case of $\epsilon(n)\propto n^{\gamma}$
$$
E_{rel}=\frac{2 \mu}{5 \gamma+2}[\frac{\gamma}{2}\frac{\dot{b}_i^2}{\omega_i^2}
+\frac{1}{\prod_jb_j^{\gamma}}],
$$
where $\mu$ is the chemical potential.

Expanding Eqs.(12) around equilibrium ($b_i=1$)   we get in the case of anisotropic trapping ($\omega_x=\omega_y=\omega_{\perp}$,
$\omega_z/\omega_{\perp}=\lambda$) the following result for the frequency of the radial compression mode
$$
\omega_{rad}=\frac{\omega_{\perp}}{\sqrt{2}} [4+2\kappa+3 \lambda^2+\kappa \lambda^2+\sqrt{(4+2\kappa+3 \lambda^2+\kappa \lambda
^2)^2-4(10+6\kappa)\lambda^2}]^{1/2},
\eqno{(14)}
$$
where 
$\kappa=\int n_0^3d^2\epsilon/(dn_0^2)d^3r/\int n_0^2d\epsilon/(d  n_0)d^3r$.
For  an elongated trap, $\lambda\ll1$, we can rewrite Eq.(14) as
$$
\omega_{rad}\approx\omega_{\perp} \sqrt{4+2\kappa}.
\eqno{(15)}
$$
 Note that Eq.(14) for the case of $\epsilon(n)\propto n^{\gamma}$
was considered in several papers [40]. 

In  Fig. 2 we present the calculations of $\omega_{rad}$ using two
 approximations, Eq.(5) and Eq.(6), for the equation of state $\epsilon(n)$
(to calculate the ground-state density we
have used a highly accurate
 variational approach of Ref.[41]).  The curves explicitly show the nonmonotonic behavior of 
$\omega_{rad}$ in the agreement with a schematic interpolation of Ref.[42].
It can be seen from Fig. 2 that the difference between two approximations,
Eq.(5) and Eq.(6), is less than 0.7\%.

 Our calculated results for the axial cloud size of strongly interacting $^6Li$ atoms as a function of the magnetic field strength
$B$ are compared with the recent experimental data [10] in Fig. 3.
This comparison shows that although both approximations, Eq.(5) and Eq.(6), 
give a reasonable agreement with experimental data, the equation of state from
 Eq.(6) leads to the better description of the experimental curve.
We have used the data from Ref.[17] to convert $a$ to $B$. 

\pagebreak

\noindent
We note here
that in general  a Feshbach resonance may lead to the density dependence of the
effective interaction
  (for bosons cases see, for example, [43,44]).

In conclusion, 
 we have considered the expansion of the cloud of initially confined $^6Li$ atoms and studied the equation of state dependence of the radii of trapped cloud and collective oscillations near the broad 
Feshbach resonance at $B=820 \pm 3$G.
It is shown a non monotonic behavior of the radial compression mode frequency 
and demonstrated that an important test of the equation of state can be 
obtained from the study of the radii of trapped cloud 
in regimes
 now available experimentally.

{\it Note added}: A recent paper by the Duke University  group [45] reports on
 measurements of the radial compression mode frequencies. Our calculations
 in a very good agreement with these experimental data  on the BCS side.

While this work was being prepared for publication, two preprints [46,47] appeared in which the authors consider collective modes and the expansion of a trapped superfluid Fermi gas in the BCS-BEC crossover. For the negative   
scattering length case, their results are in perfect agreement with ours.

\pagebreak

\begin{figure}[ht]
\includegraphics{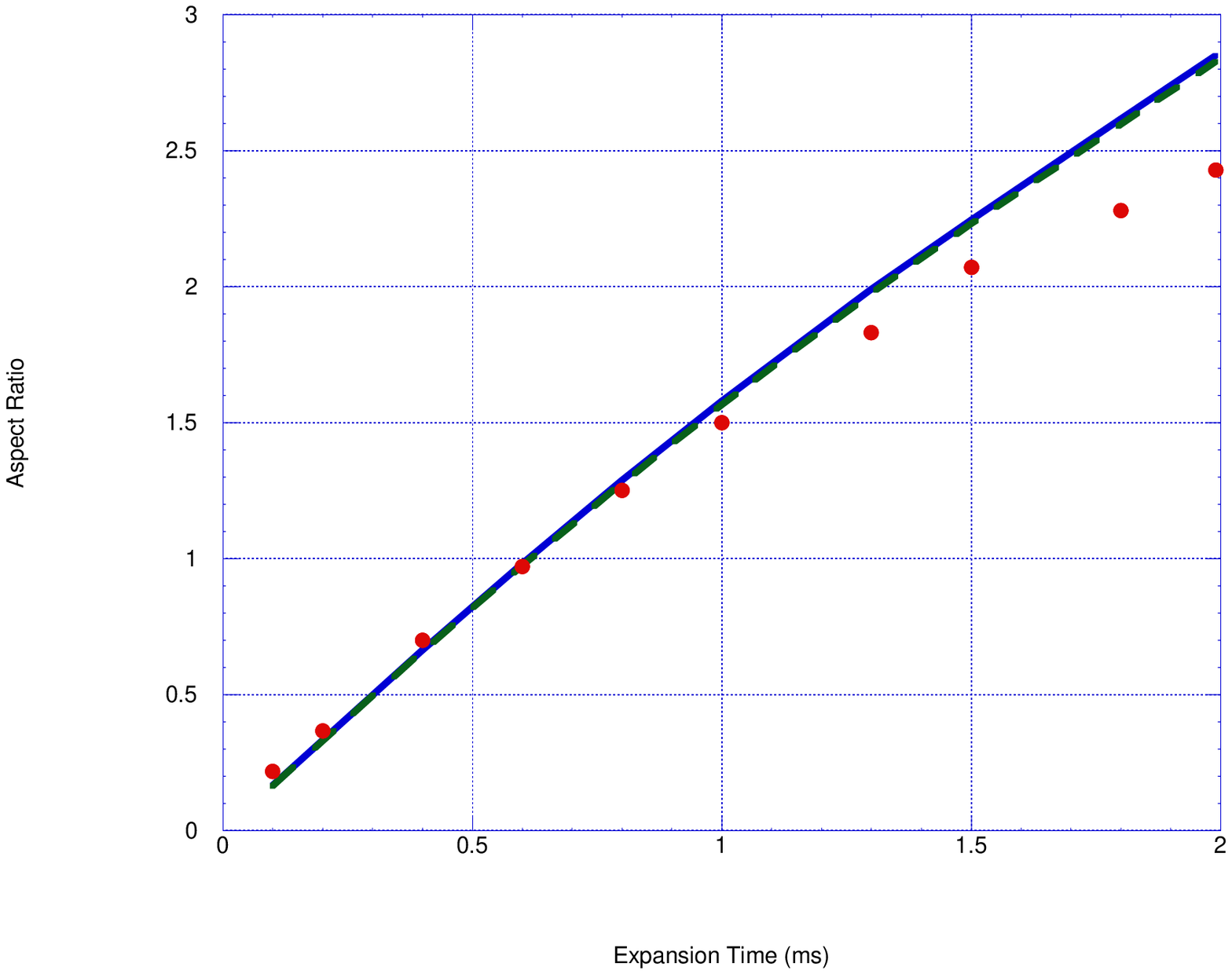}
\end{figure}
Fig. 1. Aspect ratio of the cloud of the $N=7.5 \times 10^4$ $^6Li$ atoms as a function of time
 after release from the trap ($\omega_{\perp}=2 \pi \times 6605$Hz,
$\omega_z=2 \pi \times 230$Hz).
The circular dots indicate experimental data from the Duke University group [9].
The solid line and the dashed line represent theoretical calculations in the
 unitary limit ($a\rightarrow -\infty$) including the quantum pressure term
  and in  the  hydrodynamic approximation,
 respectively.
 
\pagebreak
\begin{figure}[ht]
\includegraphics{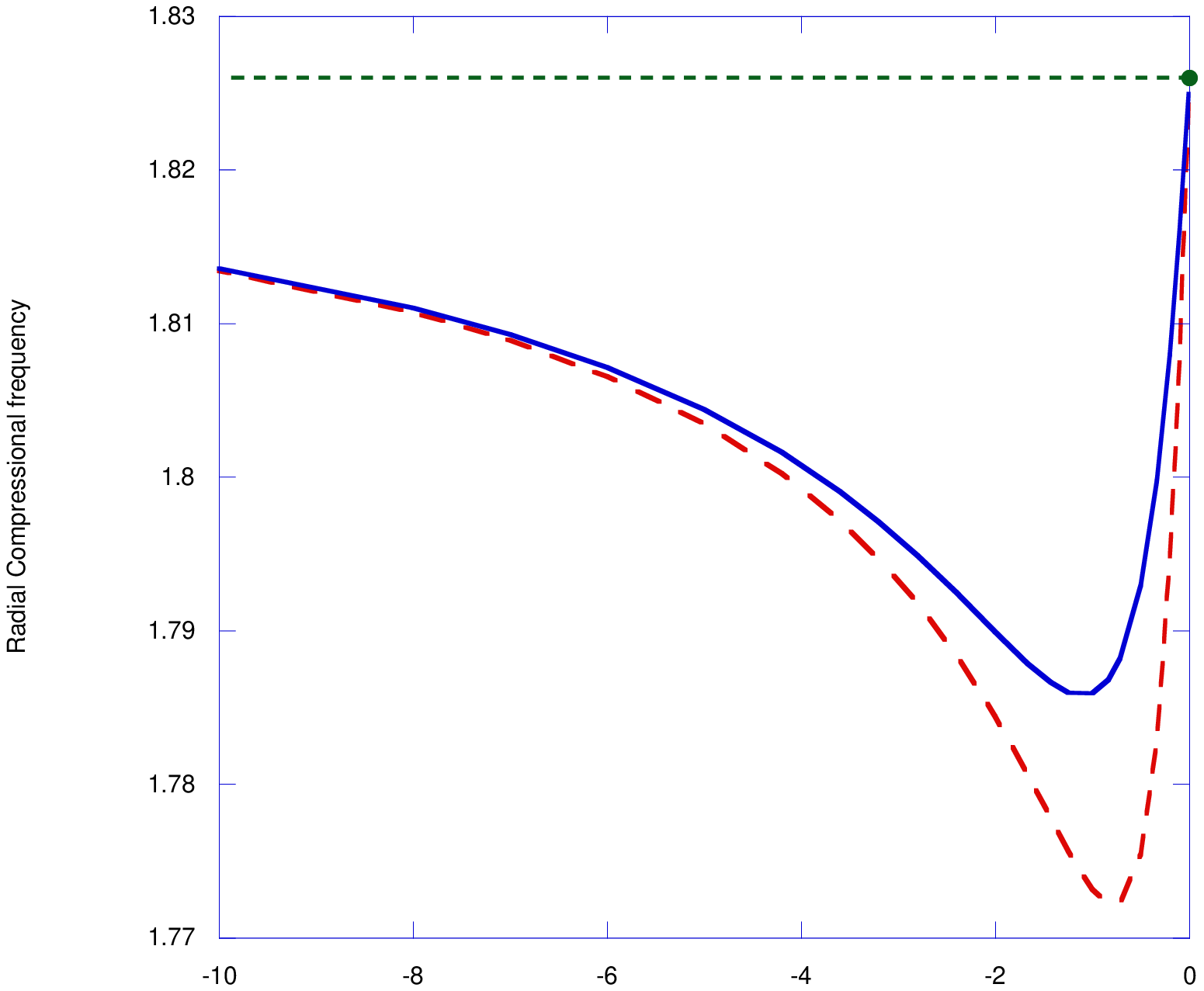}
\end{figure}
\begin{center}
$(N^{1/6}a/a_{ho})^{-1}$
\end{center}
Fig. 2. Radial compressional frequency in unit of $\omega_{\perp}$ as a function
 of the dimensional parameter $(N^{1/6} a/a_{ho})^{-1}$. In the unitary limit,
$a\rightarrow -\infty ( \bullet)$, one expect $\omega/\omega_{\perp}=\sqrt{10/3}
\approx 1.826$.
The solid line and the dashed line represent the results of theoretical
 calculations using equations of state Eq.(6) and Eq.(5) respectively.

\pagebreak

\begin{figure}[ht]
\includegraphics{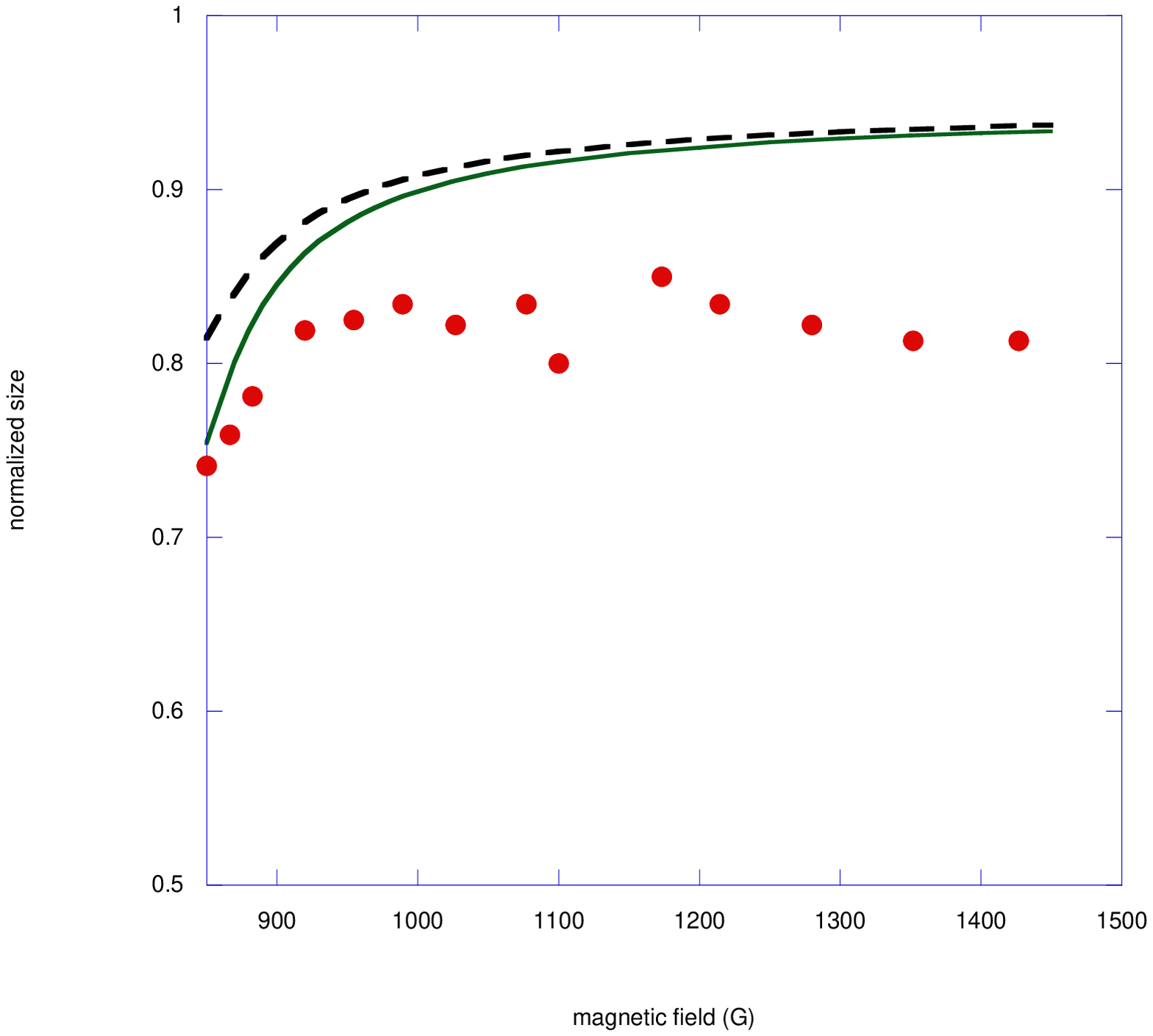}
\end{figure}
Fig. 3. Axial cloud size of strongly interacting $^6Li$ atoms
 after normalization to a non-interacting Fermi gas
with $N=4 \times 10^5$ atoms as a function of the magnetic field $B$.
 The trap parameters are $\omega_{\perp}=
2 \pi \times 640$Hz, $\omega_z=2 \pi (600 B/kG+32)^{1/2}$Hz.
 The solid line, dashed line and circular dots represent the results of
 theoretical calculations using equations of state Eq. (6), Eq. (5) and
 experimental data from the Innsbruck group [10], respectively.
\pagebreak

{\bf References}
\vspace{8pt}

\noindent
1. C.A. Regal, M. Greiner, and D.S. Jin, Phys. Rev. Lett. {\bf 92}, 040403 (2004).

\noindent
2. M. Greiner, C.A. Regal, and D.S. Jin, Nature {\bf 426}, 537 (2003).

\noindent
3. K.E. Stecker, G.B. Patridge, and R.G. Hulet, Phys. Rev. Lett. {\bf 91},
 080406 (2003).

\noindent
4. J. Cubizolles, T. Bourdel, S.J.J.M.F. Kokkelmans, G.V Shlyapnikov,
 C. Salomon, Phys. Rev. Lett. {\bf 91}, 240401 (2003).

\noindent
5. S. Jochim, M. Bartenstein, A. Altmeyer, G. Hendl, C. Chin, J.H. Denschlag, 
R. Grimm, Phys. Rev. Lett. {\bf 91}, 240402 (2003).

\noindent
6. S. Jochim, M. Bartenstein, A. Altmeyer, G. Hendl, S. Riedl, C. Chin, 
J.H. Denschlag, R. Grimm, Science {\bf 302}, 2101 (2003).

\noindent
7. M.W. Zwierlein, C.A. Stan, C.H. Schunck, S.M.F. Raupach, S. Gupta, 
Z. Hadzibabic Z, W. Ketterle, Phys. Rev. Lett. {\bf 91}, 250401 (2003).

\noindent
8. C.A. Regal and D.S. Jin, Phys. Rev. Lett. {\bf90},230404 (2003).

\noindent
9. K.M. O'Hara, S.L. Hemmer, M.E. Gehm, S.R. Granade, J.E. Thomas
Science {298}, 2179 (2002).

\noindent
10. M. Bartenstein, A. Altmeyer, S. Riedl, S. Jochim, C. Chin, J. 
Hecker Denschlag, R. Grimm, Phys. Rev. Lett. {\bf 92}, 120401 (2004).

\noindent
11. A.J. Leggett, J. Phys. (Paris), Colloq. {\bf 41}, 7 (1980).

\noindent
12. P. Nozieres and S. Schmitt-Rink, J. Low Temp. Phys. {]bf 59}, 195 (1985).

\noindent
13. M. Randeria, in {\it Bose-Einstein Condensation}, edited by A. Griffin, D.W. Snoke, and S. Stringari (Cambridge University, Cambridge, 1995),pp. 355-392.

\noindent
14.  S.J.J.M.F. Kokkelmans, M. Holland, R. Walser, M. Chiofalo,
Phys. Rev. Lett. {\bf 87},120406 (2001);
 Acta Physica Polonica
A{\bf 101}, 387 (2002).

\noindent
15. E. Timmermans, K. Furuya, P.W. Milonni, A.K. Kerman, Phys. Lett. A{\bf 285}, 228 (2001).

\noindent
16. M. Houbiers, H. T. C. Stoof, W. I. McAlexander, and R. G. Hulet,
Phys. Rev. A{\bf 57}, 1497 (1998).

\noindent
17.     K. M. O'Hara, S. L. Hemmer, S. R. Granade, M. E. Gehm, J. E. Thomas,
     V. Venturi, E. Tiesinga, and C. J. Williams, Phys. Rev. A{\bf 66},
 041401 (2002).

\noindent
18. M. W. Zwierlein, C. A. Stan, C. H. Schunck, S. M. F. Raupach,
A. J. Kerman, and W. Ketterle, Phys. Rev. Lett. {\bf 92}, 120403 (2004).

\noindent
19. 
 R. Singh and B.M. Deb, Phys. Rep. {\bf 311}, 47 (1999) and references therein.

\noindent
20. L. Pitaevskii and S. Stringari, {\it Bose-Einstein Condensation} (Clarendon Press,
Oxford 2003).

\noindent
21. C. Menotti, P. Pedri and S. Stringari, Phys. Rev. Lett. {\bf 89}, 25042 (2002).

\noindent
22. W. Vincent Liu and F. Wilczek, Phys. Rev. Lett. {\bf 90}, 047002 (2003).

\noindent
23. I. Shovkovy and M. Huang, Phys. Lett. B{\bf 564}, 205 (2003).

\noindent
24. P.F. Bedaque, H. Caldas, and G. Rupak, Phys. Rev. Lett.  {\bf 91}, 247002
 (2003).

\noindent
25. S.K. Adhikari, Am. J. Phys. {\bf 54}, 362 (1986).

\noindent
26. W. Lenz, Z. Phys. {\bf 56}, 778 (1929).

\noindent
27. K. Huang and C.N. Yang, Phys. Rev. {\bf 105},767 (1957).

\noindent
28. T.D. Lee and C.N. Yang, Phys. Rev.  {\bf 105},1119 (1957).

\noindent
29. V.N. Efimov and M.Ya. Amus'ya, Zh. Eksp. Teor. Fiz. {\bf 47}, 581 (1964) [
Sov. Phys. JETP {\bf 20},388 (1965)].

\noindent
30. G.A. Baker, Jr., Int. J. Mod. Phys. B{\bf15}, 1314 (2001) and references
 therein.

\noindent
31. H. Heiselberg, Phys. Rev. A{\bf 63}, 043606 (2001).

\noindent
32. J. Carlson, S.-Y. Chang, V.R. Pandharipande, and K.E. Schmidt, Phys. Rev. {\bf 91},
050401 (2003).

\noindent
33. Y.E. Kim and A.L. Zubarev, Phys. Rev. A{\bf 69}, 023602 (2004).

\noindent
34. Y.E. Kim and A.L. Zubarev, Phys. Rev. A{\bf 67}, 015602 (2003).

\noindent
35. H. Suno, B.D. Esry and C.H. Greene, Phys. Rev. Lett. {\bf 90}, 053202 (2003).

\noindent
36. A. Griffin, Wen-Chin Wu and S. Stringari, Phys. Rev. Lett. {\bf78}, 1838 (1996).

\noindent
37.  Yu. Kagan, E.L. Surkov and G.V. Shlyapnikov, Phys. Rev. A{\bf 55}, 18 (1997).

\noindent
38. P. Pedri, D. Gu\'{e}ry-Odelin and S. Stringari, cond-mat/0305624.

\noindent
39. A.L. Zubarev and Y.E. Kim,  Phys. Rev. A{\bf 65}, 035601 (2002).

\noindent
40.  M. Cozzini and S. Stringari, Phys. Rev. Lett. {\bf 91},
 070401 (2003)
 and references therein.

\noindent
41. M.P. Singh and A.L. Satheesha, Eur. Phys. J., D{\bf 7}, 321 (1998).

\noindent
42. S. Stringari, Europhys. Lett. {\bf 65}, 749 (2004).

\noindent
43. V.A. Yurovsky, cond-mat/0308465.

\noindent
44. Y.E. Kim and A.L. Zubarev, Phys. Lett. A{\bf 312}, 277 (2003).

\noindent
45. J. Kinast, S. L. Hemmer, M. E. Gehm, A. Turlapov, and J. E. Thomas,
 Phys. Rev. Lett. {\bf 92}, 150402 (2004).

\noindent
46. H. Heiselberg, cond-mat/0403041.

\noindent
47. Hui Hu, A. Minguzzi, Xia-Ji Liu, and M.P. Tosi, cond-mat/0404012. 
\end{document}